\documentclass[conference]{IEEEtran}
\IEEEoverridecommandlockouts

\usepackage{cite}
\usepackage{amsmath,amssymb,amsfonts}
\usepackage{algorithm}
\usepackage{algorithmic}
\usepackage{graphicx}
\usepackage{textcomp}
\usepackage{xcolor}
\usepackage{multirow}

\def\BibTeX{{\rm B\kern-.05em{\sc i\kern-.025em b}\kern-.08em
    T\kern-.1667em\lower.7ex\hbox{E}\kern-.125emX}}

\IEEEpubid{\makebox[\columnwidth]{979-8-3195-2794-3/26/\$31.00~\copyright~2026 IEEE \hfill} \hspace{\columnsep}\makebox[\columnwidth]{ }}
\begin{document}

\title{On Learning Optimal Corners in Orthogonal Partially Observable Cooperative Guard Art Galleries}

\author{
\IEEEauthorblockN{Yassin Ben Mansour}
\IEEEauthorblockA{
\textit{UFR Sciences} \\
\textit{Université Paris-Saclay}\\
Orsay, France \\
yassim.ben-mansour@universite-paris-saclay.fr}
\and
\IEEEauthorblockN{Edwin Meriaux}
\IEEEauthorblockA{
\textit{L2S, Centralesupélec} \\
\textit{Université Paris-Saclay}\\
Gif-sur-Yvette, France \\
edwin.meriaux@mail.mcgill.ca}

}

\maketitle

\begin{abstract}
The CADENCE algorithm solves the Partially Observable Cooperative Guard Art
Gallery Problem (POCGAGP) with formal coverage and connectivity guarantees,
but leaves unspecified which valid corner each agent should be deployed to,
a choice that strongly affects efficiency. We introduce two learned
corner-selection heuristics that preserve these guarantees: a CNN scoring
candidates on a grid encoding, and a GATv2 network trained with Deep
Q-Learning (DQN) on a visibility graph. Across 7{,}500 runs on random
orthogonal environments ($50^2$ to $250^2$), our heuristics
outperform baseline CADENCE in both steps to full coverage and peak agent
count, with gains growing with scale, and improve on Incremental
Self-Deployment (ISDA) baselines in agent utilization while providing
guarantees ISDA lacks. Learned corner selection thus improves CADENCE
 in speed and agent utilization at no cost to its formal properties.
\end{abstract}


\section{Introduction}

\begin{figure}[h!]
    \centering
    \includegraphics[width=0.8\linewidth]{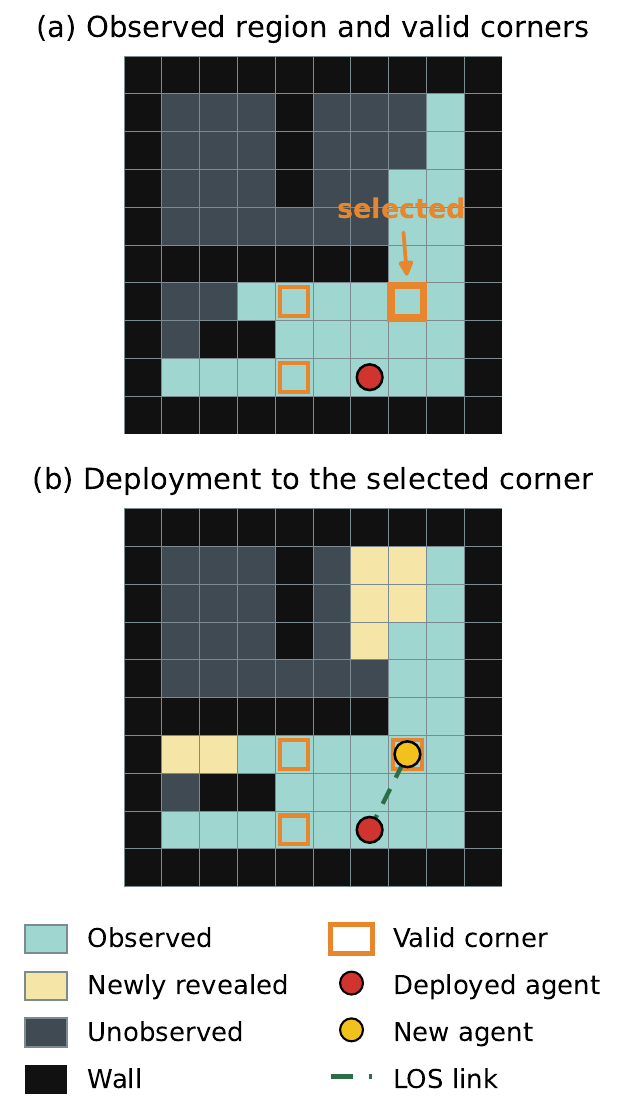}
    \caption{Agent Deployment in CADENCE, previous agent (red color), new agent deployed in a valid corner (yellow color)}
    \label{fig:placeholder}
\end{figure}

Deploying mobile robot teams in unknown environments for disaster response and inspection requires balancing two coupled challenges: exploring the area and maintaining a mobile ad hoc network (MANET) for communication. Because new observations alter the topology and the map is unknown in advance, this dual requirement cannot be solved through offline planning.
 
The computational geometry community represents exactly this situation as
the Partially Observable Cooperative Guard Art Gallery Problem (POCGAGP).
POCGAGP is the partially observable variant of the Cooperative Guard Art
Gallery Problem (CGAGP), which in turn extends the classical Art Gallery
Problem (AGP)~\cite{o1987art}. AGP asks for the smallest set of static guards whose
visibility polygons cover a given polygon, and computing that minimum set is
NP-hard~\cite{1057165}. CGAGP
adds the requirement that the guards form a connected visibility graph,
which is precisely the Mobile Ad Hoc Network (MANET) condition once mutual visibility is read as a
line-of-sight link. POCGAGP then removes the assumption that the polygon is
known in advance, leaving agents to extend coverage and connectivity from
whatever they have observed so far.
 
CADENCE solves POCGAGP by deploying agents one at a time to valid corners on the boundary of the observed region, this makes the number of deployed agents bounded by the number of valid corners: 
$M_{\text{valid}} := \frac{n + 2h - 4}{2}$, where n is the number of $90^\circ$ or $270^\circ$ interior angles, and h is the number of holes~\cite{meriaux2025mobile}. A valid corner~\cite{meriaux2025mobile} is a 
$270^\circ$ reflex corner which is not at the top left of a hole (an obstacle not part of the outside wall) within the environment.

Reflex corners form the ideal candidate set as they mark the exact vertices where visibility is occluded, resolving line-of-sight barriers no interior placement can. This formulation is strictly two-dimensional and orthogonal: both the guard bounds and discrete corner extraction rely on orthogonality, surviving neither a move to general polygons nor to 3D. Since our aim is to improve CADENCE rather than propose a new solver, we operate directly within its defined setting.

While CADENCE identifies candidate corners, it lacks a criterion for selecting among them. These decisions compound over time because early placements dictate future candidate availability; a myopically attractive choice can easily preclude a shorter overall solution. However, evaluating candidates explicitly requires exhaustive visibility simulation and downstream lookahead—the exact computational burden the incremental formulation aims to avoid.

We replace this missing criterion with a learned scoring function that predicts candidate values and selects the optimal deployment. The central question is state representation: grid-based encodings allow convolutional networks to exploit spatial structure, whereas graph representations capture relational visibility between corners and cells. Comparing these distinct encodings forms the substance of this work. Our contributions are:
 
\begin{itemize}
  \item two learned valid-corner selection heuristics for CADENCE: a dense
        CNN over a grid encoding, GATv2 networks over a
        visibility-graph encoding;
  \item an evaluation against
        CADENCE $rand\_point$ selection, ISDA edge and ISDA any
        on generated orthogonal environments, measuring agents deployed and steps to full coverage. Overall 7,500 test were run to validate the findings.
\end{itemize}
 
\section{Background}
 
\subsection{POCGAGP and CADENCE}
 
Classical art gallery results assume the polygon is given in advance. The
robotics exploration literature relaxes that assumption using frontier-based
methods that drive robots toward the boundary between known and unknown
space~\cite{yamauchi1997frontier}, but those methods are formulated over occupancy grids and carry
neither a geometric coverage guarantee nor an explicit connectivity
constraint. POCGAGP bridges the two, and CADENCE realizes it by maintaining
the observed region, extracting its valid $270^\circ$ corners, deploying an
agent to one of them, and repeating until coverage is complete~\cite{meriaux2025mobile}. We modify
neither the POCGAGP formulation nor the connectivity machinery; we target
only the unspecified selection inside CADENCE's loop.
 
\subsection{Grid and Graph Representations}
 
Convolutional networks exploit translation equivariance and local receptive
fields~\cite{lecun1998gradient}, and an orthogonal environment discretizes onto a grid without
distortion. The quantities determining a corner's value are local spatial
patterns: unseen area beyond it, surrounding wall orientation, and frontier
position. This makes a convolutional encoder a natural first hypothesis.
Its limitation is that visibility is not a spatially local relation: a corner
can reveal a cell far along a sightline while being blind to its immediate
neighbour behind a wall, so convolution must be deep enough to span
sightlines and must additionally learn occlusion logic the geometry already
provides.
 
Graph attention networks~\cite{velickovic2018graph} operate directly on the visibility relation itself. When edges encode sightlines rather than spatial adjacency, a single message-passing step immediately propagates coverage information along line-of-sight paths. We use the GATv2 formulation, which resolves a key limitation of standard graph attention by allowing node rankings to depend on the target node. This distinction is critical here because visible cells vary in value: an already-observed cell adds nothing, whereas an unexplored frontier cell is highly valuable.
 
\subsection{Learning to Score Candidates}
 
Guiding a combinatorial search with a learned scoring function is an
established pattern in which the search retains its correctness conditions
and learning supplies only the ordering over candidates. A mis-scored
candidate therefore costs efficiency rather than correctness, since CADENCE
terminates with full coverage regardless of which valid corner is chosen.
Because a placement's value depends on the placements that follow it,
selection is naturally sequential; we train the graph-based selectors by
deep Q-learning~\cite{mnih2015human} against the CADENCE rollout, using newly discovered nodes as
the reward signal.

\section{Problem Formulation}
The central problem studied in this work is choosing the optimal valid corner at each step of CADENCE to achieve full coverage of the world while minimizing both the total time steps required and the number of deployed agents. The core decision mechanism at each decision epoch relies on selecting an unvisited, valid candidate location: $x \in S'_k \setminus (T_k \cup \{x_d\})$.
 With $S'_k$ is the set of valid candidate corners evaluated at step $k$, $T_k$ is the set of already visited or covered target locations up to step $k$ and $x_d$ is the deployment point.

Our objective is to learn or compute an optimal policy $\pi^*$, parameterized by our heuristic/model, that selects the best target point $x^* \in  S'_k \setminus (T_k \cup \{x_d\})$.

$$x^* = \underset{x \in S'_k \setminus (T_k \cup \{x_d\})}{\operatorname{argmin}} C(S_k, x; \theta)$$
Where $C(S_k, x; \theta)$ quantifies the estimated remaining cost (in time steps and potential agent additions) to complete coverage starting from target choice $x$, and $S_k$ is the environment state at step k.

\section{Algorithms}

We now present two learning-based algorithms that will reduce the overall cost for CADENCE of solving the POCGAGP while maintaining the known $\frac{n}{2} +h -2$ bounds as they only select from valid corners. 

\subsection{Convolutional Neural Network Method}

\begin{algorithm}[t]
\caption{CNN-based Corner Selection for CADENCE}
\label{alg:cnn_cadence}
\begin{algorithmic}[1]
\STATE \textbf{Input:} Environment state $S_0$
\STATE \textbf{Output:} Agent positions $\mathcal{A}$
\WHILE{world not fully covered}
    \STATE Extract valid $270^\circ$ corners $S'_k$
    \STATE $(x^*, y^*) \gets \arg\max_{(x,y) \in S'_k} f_\theta(S_k, x, y)$
    \STATE Deploy agent at $(x^*, y^*)$; add to $\mathcal{A}$
    \STATE Update observed region
    \STATE De-allocate redundant agents
\ENDWHILE\\
\textbf{RETURN} $\mathcal{A}$
\end{algorithmic}
\end{algorithm}

\begin{figure}
    \centering
    \includegraphics[width=0.95\linewidth]{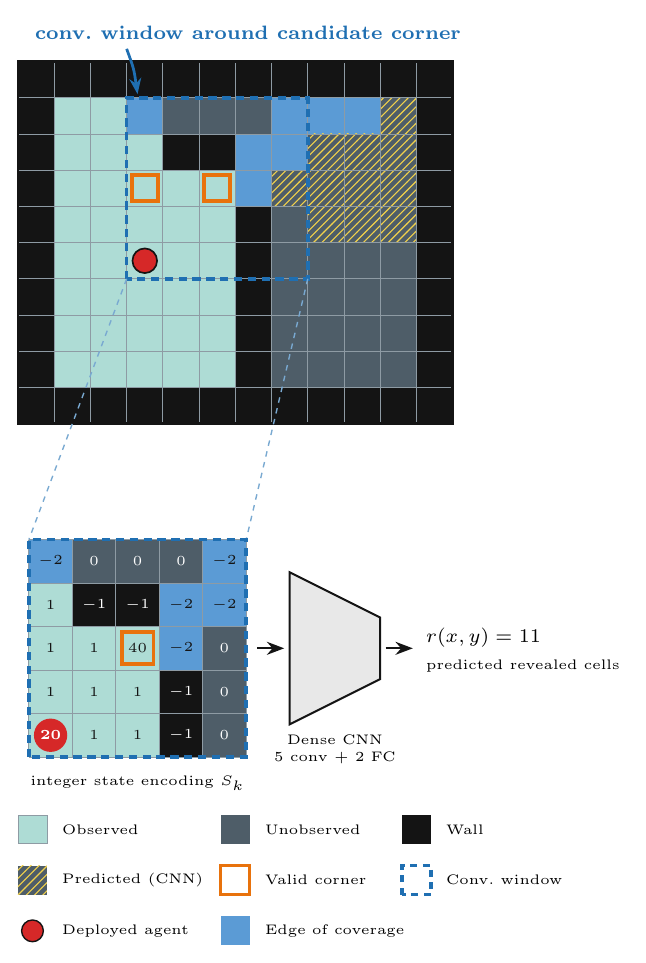}
    \caption{CNN-based corner selection, agent (red color), valid corners (orange color), new agent(green color), predicted discovered area(yellow color)}
    \label{fig:cnn}
\end{figure}

\begin{figure}
    \centering
    \includegraphics[width=1\linewidth]{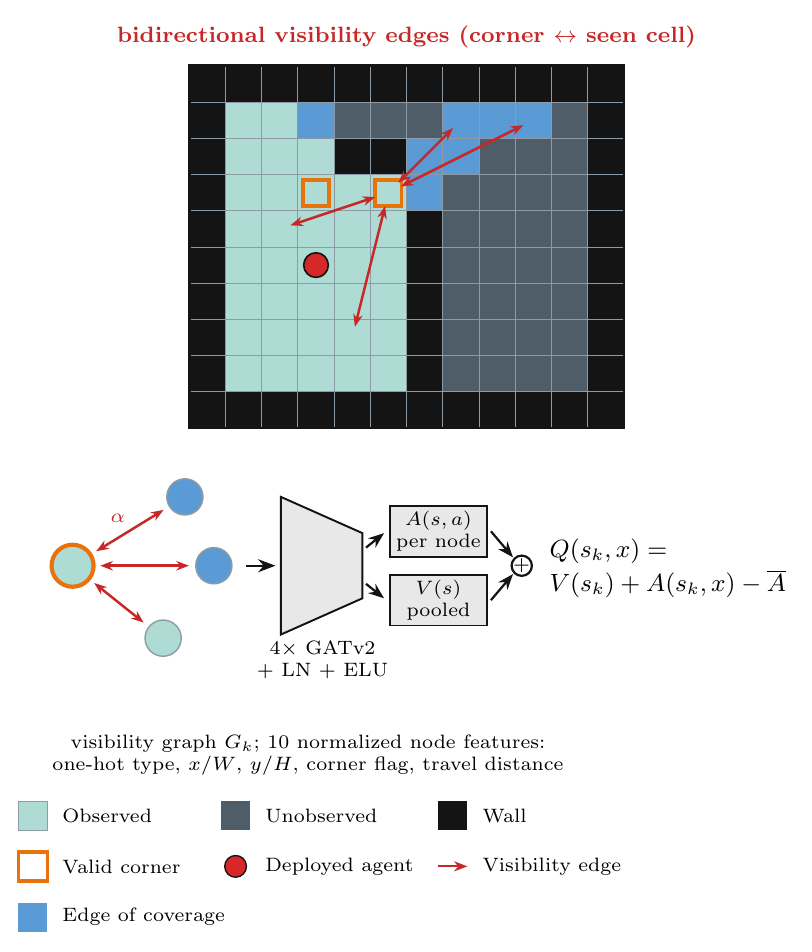}
    \caption{GATv2-based scoring of a candidate valid corner. }
    \label{fig:gat_scoring}
\end{figure}

\subsubsection{State Representation}
The environment state $S_k$ is an $H \times W$ integer matrix in which each
entry encodes the status of one grid position: $-2$ for edge of seen area, $-1$ for seen walls, $0$ for undiscovered area, $1$ for free seen space, $20$ for agents, $30$ for the deployment point and $40$ for the new agent (used to predict discovered area).

\subsubsection{Network Architecture}
The scoring function $f_\theta$ is a dense CNN with five convolutional layers
and two fully connected layers, using ReLU activations, batch normalization
(to stabilize training), and average pooling. Given the state $S_k$ and a
candidate corner $(x,y)$, it produces a single output: the expected number
of cells revealed by deploying an agent there.

\subsubsection{Integration into CADENCE}
Algorithm~\ref{alg:cnn_cadence} shows the resulting loop. At each step, the
valid $270^\circ$ corners are extracted from the observed region, $f_\theta$
scores each candidate, and the agent is deployed to the candidate that maximizes $f_\theta$. As shown
in Fig.~\ref{fig:cnn}, the network evaluates a window around each
valid corner (orange) and predicts the area that would be revealed (yellow);
the corner with the highest predicted area is selected. Only the selection
among valid corners changes; deployment, observation updates, and
de-allocation follow CADENCE unchanged.

\subsubsection{Training Pipeline}
Training data was generated locally as pairs of environment states before
and after placing an agent at a given position, so the ground-truth label
for each sample is the number of cells that placement actually revealed.
We trained the CNN on an augmented dataset of over 100 pairs with 100 epochs.
The CNN is trained to predict this count. Our working assumption is that
greedily deploying to the corner with the highest predicted revealed area
will, over the run, reduce the total time steps and agents needed for full
coverage.

\begin{algorithm}[t]
\caption{DQN-based Corner Selection for CADENCE}
\label{alg:dqn_cadence}
\begin{algorithmic}[1]
\STATE \textbf{Input:} Environment state $S_0$
\STATE \textbf{Output:} Agent positions $\mathcal{A}$
\WHILE{world not fully covered}
    \STATE Extract valid $270^\circ$ corners $S'_k$
    \STATE Build visibility graph $G_k$ from observed region
    \STATE $(x^*, y^*) \gets \arg\max_{(x,y) \in S'_k} Q_\phi(G_k, x, y)$
    \STATE Deploy agent at $(x^*, y^*)$; add to $\mathcal{A}$
    \STATE Update observed region
    \STATE De-allocate redundant agents
\ENDWHILE \\
\textbf{RETURN} $\mathcal{A}$
\end{algorithmic}
\end{algorithm}

\subsection{Graph Attention Method}

\subsubsection{State Representation}
The visibility data is converted into a graph $G_k$ with $H \times W$ nodes,
one per position in the orthogonal environment. Edges connect each valid
$270^\circ$ corner to the positions it sees, in both directions, so that a
corner also aggregates information about the region it would reveal. Each node
carries ten normalized features: a one-hot encoding of the node type
(unexplored, free seen space, wall adjacent to seen area, frontier, agent,
deployment point), the $x$ and $y$ coordinates divided by the grid extent, a
flag marking valid corners, and the travel distance from the deployment point.

\subsubsection{Network Architecture}
The Q-network $Q_\phi$ is a graph attention network composed of four
GATv2Conv layers, each followed by layer normalization and an ELU activation,
including a residual connection between the first two layers. The resulting
node embeddings feed two parallel, two-layer MLPs: an Advantage head applied
per node, estimating action advantages $A(s, a)$, and a Value head applied to a
global mean-pooled graph embedding, estimating the state value $V(s)$. The
final action-value function $Q(s, a)$ is constructed by aggregating both
streams with mean-centering on the advantage predictions. Because the
advantage stream emits one scalar per node rather than a fixed-width vector,
the same weights apply to environments of any size. Given $G_k$, it outputs
the estimated Q-values of deploying agents in each node.

\subsubsection{Integration into CADENCE}
Algorithm~\ref{alg:dqn_cadence} shows the resulting loop. At each step, the
visibility graph $G_k$ is built from the observed region, $Q_\phi$ scores
all nodes, and the agent is deployed to the maximizer that is an element of $S'_k$, as
illustrated in Fig.~\ref{fig:gat_scoring}. As with the CNN variant, only
the selection among valid corners changes; the rest of CADENCE is
unchanged.

\subsubsection{Training Pipeline}
Unlike the CNN, which is trained on precomputed before/after pairs, the
GAT is trained directly on CADENCE rollouts by Deep
Q-Learning~\cite{mnih2015human}, with the number of newly discovered nodes
in each episode as the reward signal. We used 100 worlds, training the model on each world eight times. The working assumption is the same:
selecting the corner that reveals the most nodes should, over the run,
reduce the total time steps and agents needed for full coverage.

\section{Methodology}
The tests consist of achieving coverage for partially observed orthogonal worlds.
Experiments are conducted over randomly generated dungeon environments with sizes of 50$\times$50, 100$\times$100, 250$\times$250 (same as in previous POCGAGP benchmarks\cite{meriaux2025mobile}). We compared the CADENCE variants with our integrated CNN and DQN alongside the basic CADENCE with no policy for choosing the optimal valid corner called CADENCE $rand\_point$, and also the Incremental Self-Deployment Algorithm (ISDA) with its two variants: ISDA edge randomly selects a frontier cell, while ISDA any randomly selects any
visible cell.
Performance is evaluated by comparing the mean values of the total number of steps, final number of agents and the maximal number of agents needed to achieve full coverage of the environment, the top performing algorithm is the one that succeeds at achieving full coverage with lower values.

\section{results}

To evaluate the efficiency and scalability of our proposed CNN- and DQN-based corner selection methods in CADENCE, we compared them to CADENCE with random corner selection and two ISDA variants (ISDA edge and ISDA any) across random orthogonal worlds of varying grid sizes ($50^2$, $100^2$, and $250^2$). Performance was measured using three key metrics: the mean final number of agents, the maximal number of agents during execution, and the total number of steps taken. Lower values indicate superior performance across all metrics. Overall, for each algorithm, 300 worlds were tested, each one five times (1,500 tests per algorithm, 7,500 total). 

The CNN- and DQN-based approaches demonstrated a superior balance between execution speed and peak agent management, particularly as the environment scale increased. On small maps ($50^2$ size), CADENCE CNN and CADENCE DQN achieved full coverage in just $751$ and $816$ steps respectively, outperforming the CADENCE $rand\_point$ (1488 steps) while remaining competitive with ISDA edge (573). For larger worlds of sizes $100^2$ and especially $250^2$, the CNN's and DQN's efficiency becomes more visible, with 17,191 and 19,000 steps respectively in $250^2$ environments, compared to 51,434 for CADENCE $rand\_point$. Crucially, CADENCE DQN consistently achieved the lowest peak agent expansion across medium and large environments, recording the overall lowest maximum agent counts on $100^2$ ($31.2$) and $250^2$ ($53.2$), outperforming both $\text{rand\_point}$ ($74.2$) and $\text{ISDA edge}$ ($63.1$). While $\text{rand\_point}$ maintained slightly lower final agent counts ($44.6$ vs. $47.2$ for DQN and $48.8$ for CNN on $250^2$), both learning-based CADENCE variants consistently outperformed ISDA variants in final agent minimization, all while remaining completely resilient to the state-space explosion that caused $\text{ISDA any}$ to time out or exceed the step budget on larger maps.

\begin{table*}[t]
\centering
\caption{Mean final nb. of agents (Final), maximal nb. of agents (Max), and total nb. of steps (Steps) over random orthogonal worlds (lower is better; bests per column in bold).
Dashes indicate less than 80\% of the runs finished within the step budget for $100^2$, and 0\% of the runs finished in under 24 hours for $250^2$. \cite{meriaux2026control}}
\label{tab:results_combined}
\setlength{\tabcolsep}{3pt}

\resizebox{\textwidth}{!}{%
\begin{tabular}{lcccccccccccccccccc}
\hline
 & \multicolumn{6}{c}{$50^2$} & \multicolumn{6}{c}{$100^2$} & \multicolumn{6}{c}{$250^2$} \\
\cline{2-7}\cline{8-13}\cline{14-19}
Method & Final & $\sigma$ & Max & $\sigma$ & Steps & $\sigma$ & Final & $\sigma$ & Max & $\sigma$ & Steps & $\sigma$ & Final & $\sigma$ & Max & $\sigma$ & Steps & $\sigma$ \\
\hline
\multicolumn{19}{l}{\textit{CADENCE}} \\
\hline
rand\_point & $\mathbf{13.1}$ & $3.2$ & $22.2$ & $3.0$ & $1488$ & $412$ & $\mathbf{25.7}$ & $3.6$ & $42.3$ & $4.3$ & $8320$ & $1345$ & $\mathbf{44.6}$ & $6.3$ & $74.2$ & $6.9$ & $51434$ & $6576$ \\
CNN & $14.7$ & $3.1$ & $17.0$ & $3.5$ & $751$ & $317$ & $28.2$ & $3.5$ & $34.7$ & $4.4$ & $3552$ & $1092$ & $48.8$ & $6.2$ & $61.2$ & $7.4$ & $17191$ & $3765$ \\
DQN & $14.2$ & $3.1$ & $16.6$ & $3.2$ & $816$ & $295$ & $27.3$ & $3.5$ & $\mathbf{31.2}$ & $3.5$ & $3689$ & $937$ & $47.2$ & $6.1$ & $\mathbf{53.2}$ & $6.3$ & $19000$ & $3636$ \\
\hline
\multicolumn{19}{l}{\textit{Baseline}} \\
\hline
ISDA edge & $16.1$ & $3.3$ & $\mathbf{16.3}$ & $3.3$ & $\mathbf{573}$ & $247$ & $33.4$ & $4.0$ & $33.6$ & $4.1$ & $\mathbf{2547}$ & $800$ & $62.9$ & $6.8$ & $63.1$ & $6.8$ & $\mathbf{11561}$ & $2691$ \\
ISDA any & $13.5$ & $3.0$ & $30.2$ & $7.8$ & $6120$ & $3945$ & -- & -- & -- & -- & -- & -- & -- & -- & -- & -- & -- & -- \\
\hline
\end{tabular}%
}
\end{table*}

\setlength{\textfloatsep}{5pt}
\setlength{\intextsep}{5pt}

\section{Discussion}
\label{sec:discussion}

Table~\ref{tab:results_combined} shows that no method dominates: each of the
three metrics has a different winner, DQN achieves the best per-column score twice—winning peak agent suppression (\text{Max}) on both $100^2$ (31.2) and $250^2$ (53.2). Thus, the core contribution of the learned selectors (CNN and DQN) is a fundamental repositioning of CADENCE along its speed–resource trade-off curve rather than a simple leaderboard victory.

Against random corner selection, both methods roughly halve completion time
at $50^2$ and the advantage grows with scale, saving up to $67\%$ (CNN) and $63\%$ (DQN) of steps on $250^2$ maps. DQN excels at reducing the peak agent count by $28.3\%$ on $250^2$ maps compared to $rand\_point$. Both models pay for this with a modest increase in final agents
($6$--$12\%$), a gap that remains within one within-column standard
deviation at every size. The learned policies, trained to maximize newly
revealed area, thus concentrate their benefit on the transient cost of
the deployment---time and peak concurrency---rather than on the size of
the terminal configuration.

The comparison with ISDA edge exposes the opposite trade-off. ISDA edge
remains the fastest method at every size, and its advantage in steps
widens from $+31\%$ to $+49\%$ as worlds grow. However, DQN beats ISDA edge in peak agent count throughout at $100^2$ ($31.2$ vs. $33.6$) and $250^2$ ($53.2$ vs. $63.1$). While both DQN and CNN achieve a lower final agent count ($47.2$ and $48.8$ respectively) than ISDA edge ($62.9$) at $250^2$, representing a $22.4\%$ and $24.9\%$ reduction.

Structurally, the learned selectors merely reorder corner evaluations within CADENCE without modifying its core deployment logic. They fully retain CADENCE's worst-case agent bound ($\frac{n}{2}+h-2$) and theoretical coverage guarantees. In contrast, ISDA lacks formal safety certificates, leading to catastrophic failure where ISDA any timed out on all $250^2$.

In summary, while CNN and DQN are slower than ISDA in terms of steps, they match or exceed it in terms of agent utilization while maintaining the formal coverage and connectivity guarantees required for MANETs.
\vspace{-2mm}
\section{conclusion}
This study shows how grid-based (with Convolutional Neural Networks: CNN) and graph-based (with Graph Attention Networks (GAT) trained by Deep Q-Learning: DQN) methods can be used to optimize valid corner selection in CADENCE in order to minimize both the number of steps and agents needed to achieve full coverage of orthogonal environments in mobile ad hoc networks (MANETs) under the Partially Observable Cooperative Guard Art Gallery Problem (POCGAGP). The results obtained show how both CADENCE CNN and CADENCE DQN variants significantly outperform CADENCE with random corner selection, achieving up to $67\%$ reduction in the number of steps while maintaining strict bounds on the maximal number of agents (up to $28.3\%$ reduction). Crucially, while traditional baselines like ISDA any suffer severe scalability bottlenecks and fail on large $100^2$ and $250^2$ environments, CADENCE CNN and DQN maintain consistent execution efficiency and stability in large grids. These results highlight CNN- and DQN-driven heuristics as a practical, scalable solution for multi-agent coverage tasks in mobile ad hoc networks where agent overhead and execution time are heavily constrained. Future work will extend this framework to non-orthogonal worlds and explore graph attention mechanisms with focus on global reward to further optimize its performance.

\vspace{-2mm}
\bibliographystyle{IEEEtran}
\bibliography{references}

\end{document}